\documentclass{article}
\usepackage{epsfig}
\usepackage{amsmath, amsthm, amsfonts}

\newcommand{\PP}{{\mathbb P}}
\newcommand{\R}{{\mathbb R}}
\newcommand{\C}{{\mathbb C}}

\newcommand{\Prob}{{\rm Prob}}

\renewcommand{\phi}{\varphi}

\newtheorem{theo}{{\sc Theorem}}[section]

\newtheorem{prop}[theo]{{\sc Proposition}}

\begin{document}

\sc
\title{$SU(1,1)$ Random Polynomials}
\author{Pavel Bleher, Denis Ridzal \\ \\ 
\small Department of Mathematical Sciences \\ \small 
Indiana University -- Purdue University Indianapolis \\ \small
Indianapolis, IN  46202-5132\\}
\maketitle

\rm

\begin{abstract}
We study statistical properties of zeros of random polynomials 
and random analytic functions associated with the
pseudoeuclidean group of symmetries $SU(1,1)$,  
by utilizing both analytical and numerical techniques. We first show
that zeros of the $SU(1,1)$ random polynomial of degree $N$
are concentrated in a narrow annulus of the order of $N^{-1}$
around the unit circle on the complex plane, and we find an explicit formula 
for the scaled density of the zeros distribution along the radius
in the limit $N\to\infty$.
 Our results are supported through 
various numerical simulations. We then extend results of Hannay \cite{H}
and Bleher, Shiffman,
Zelditch \cite{BSZ2} to derive different formulae for
correlations between zeros of the $SU(1,1)$ 
random  analytic functions, by applying the generalized Kac-Rice formula.
We express the correlation functions in terms of some Gaussian integrals,
which can be evaluated combinatorially as a finite sum over Feynman
diagrams or as a supersymmetric integral.   
Due to the $SU(1,1)$ symmetry, the correlation functions depend
only on the hyperbolic distances between the points on the unit disk, 
and we obtain an explicit 
formula for the two point correlation function. It displays quadratic
repulsion at small distances and fast decay of correlations at infinity. 
In an appendix
to the paper we evaluate correlations between the outer zeros $|z_j|>1$
of the $SU(1,1)$ random polynomial, and we prove that the inner
and outer zeros are independent in the limit when the degree of
the polynomial goes to infinity.
\\
\end{abstract}

\section{Introduction}

In this paper we are interested in  statistical properties of zeros of
random polynomials and random analytic functions associated with the
pseudoeuclidean group of symmetries $SU(1,1)$. The motivation for the study of zeros
of random polynomials and random analytic functions comes from
different applications, most importantly from the theory of quantum chaos
(see papers of Bogomolny, Bohigas, and Leboeuf \cite{BBL},
Leboeuf and Shukla \cite{LS}, 
 Hannay \cite{H}, Korsch, Miller, and  Wiescher \cite{KMW},  
Nonnenmacher and Voros \cite{NV}, Forrester and
Honner \cite{FH}, Leboeuf \cite{L}, Shiffman and Zelditch \cite{SZ},
and others).
There are different ensembles of random polynomials associated
with different groups of symmetries. In this respect we can make a
parallel to the theory of random matrices. In the latter, the
basic ensembles include orthogonal, unitary, symplectic,
circular, and some others (see \cite{M}). Making the parallel, we may think of
the $O(n+1)$ ensemble of real polynomials as of an analogue of
the orthogonal ensemble of random matrices. The $O(n+1)$
ensemble of random polynomials consists of multivariate homogeneous real
polynomials of
$(n+1)$ variable $z~=~(z_0,z_1,\dots,z_n)$ of the form
\begin{equation}\label{ON}
\psi(z) = \sum_{|m|= N}{\sqrt{C_{N}^{m}}\;a_{m}z^{m}}\mbox{,}
\end{equation}
where $m=(m_0,m_1,\dots,m_n)$ is a multiindex, $|m|=m_0+m_1+\dots+m_n$,
$z^m=z_0^{m_0}z_1^{m_1}\dots z_n^{m_n}$,
$C_N^m$ is the multinomial coefficient,
\begin{equation}\label{CNm}
C_{N}^{m}\equiv\frac{N!}{m_0!m_1!\dots m_n!}\mbox{,}
\end{equation}
and $a_m$ are independent standard, $N(0,1)$, real Gaussian random variables.
Consider a set $Z$ of joint real zeros of $k$, $k\le n$, independent copies
of the $O(n+1)$ random polynomial in the projective space $\R\PP^n$.
 This is a random real algebraic
variety of dimension $n-k$. It is nondegenerate almost surely
and it possesses a natural volume element induced by the
standard metric in $\R\PP^n$. The joint distribution functions
of the zeros are invariant with respect to the action of the group
$O(n+1)$, and, in particular, the density function of the zeros
is constant (cf. \cite{ShSm}, \cite{BSZ2}, \cite{BSZ3}). 

Similarly, the $SU(n+1)$ ensemble of random polynomials
consists of multivariate homogeneous complex
polynomials of
$(n+1)$ complex variable $z=(z_0,z_1,\dots,z_n)$ of form (\ref{ON}) where
$a_m$ are independent standard complex Gaussian random variables
(cf. \cite{BBL}, \cite{H}, \cite{BSZ1}-\cite{BSZ4}). It corresponds
to the unitary ensemble of random matrices. The distribution
of joint zeros of $k$ independent copies of the $SU(n+1)$
random polynomial is invariant with respect to the action
of the group $SU(n+1)$ on the complex projective space $\C\PP^n$.
As the degree $N$ of the $SU(n+1)$ random polynomial goes to
infinity, the scaled correlation functions of zeros approach
a limit, which is represented by the correlation functions
of the $W_n$ ensemble of random analytic functions
(see \cite{H}, \cite{BSZ1}-\cite{BSZ4}, \cite{L}). The latter
consists of random functions of the form
\begin{equation}\label{Wn}
\psi(z) = \sum_{m}\sqrt{\frac{1}{m!}}\;a_{m}z^{m}\mbox{,}
\end{equation}
where the sum runs over multiindices $m=(m_1,\dots,m_n)$ with 
$m_j\ge 0$, $m!=m_1!\dots,m_n!$, and 
$a_m$ are independent standard complex Gaussian random variables.
The existence of the limit for the scaled correlation functions 
of zeros is valid
in a very general framework of random sections of powers of
line bundles over compact manifolds and the limit is universal
(see \cite{BSZ1}-\cite{BSZ4}). The explicit combinatorial formulae for
the limit of two-point correlation functions are obtained in
\cite{BSZ2}-\cite{BSZ4}. They demonstrate a quadratic repulsion
if $k=n=1$, neutrality if $k=n=2$, and attraction if $k=n>2$.

In this paper we are interested in the pseudoeuclidean, $SU(1,1)$ 
ensemble of random analytic functions. The general SU$(n,1)$
ensemble consists of multivariate analytic functions of the form
\begin{equation}\label{SUn1}
\psi(z) = \sum_{m}{\sqrt{C_{|m|+L-1}^{m}}\;a_{m}z^{m}}\mbox{,}
\qquad z=(z_1,\dots,z_n) \mbox{,}
\end{equation}
where $L\ge 1$ is a fixed integer, a parameter of the ensemble,
the sum runs over multiindices $m=(m_1,\dots,m_n)$ with 
$m_j\ge 0$, $z^m=z_1^{m_1}\dots z_n^{m_n}$, 
\begin{equation}\label{CNmL}
C_{|m|+L-1}^{m}\equiv\frac{(|m|+L-1)!}{(L-1)!m_1!\dots m_n!}\mbox{,}
\end{equation}
and
$a_m$ are independent standard complex Gaussian random variables.
We will restrict our study to the case $n=1$.
The basic calculations are extended to the case of any $n$, and
we are going to return to this extension in
subsequent publications. Another possible extension concerns
the universality result like in \cite{BSZ2}, \cite{SZ2}, and \cite{BSZ3}. 

The plan of the paper is the following.  In Section \ref{S:basic} we
consider the ensemble of $SU(1,1)$ random polynomials, which
is obtained by restricting $m$ in (\ref{SUn1}) to be bounded by $N$
(with $n=1$).
We calculate the density function for the distribution of zeros
of the $SU(1,1)$ random polynomial and we show that the zeros
are concentrated in a narrow annulus around the unit circle,
of the width of the order of $1/N$. We find the scaled
profile of the density function along the radius in
the limit $N\to\infty$. In Subsection
\ref{SS:SU11} we rederive the result of Leboeuf \cite{L} for
the density of the $SU(1,1)$ random analytic function.
In Section \ref{S:corr} we derive different formulae for the
correlation functions between zeros of the $SU(1,1)$ random analytic function.
First, we apply the general result of \cite{BSZ2} to get the Kac-Rice type
expression for the $n$-point  correlation function. We then
specify it for the case of a Gaussian random field (cf. \cite{H},
\cite{BD}, \cite{BSZ2}).
This expresses the $n$-point  correlation function in terms of
some Gaussian averages, which is further represented as a sum
over Feynman diagrams, or in a different approach developed in
\cite{BSZ4}, as a supersymmetric integral. In Subsection
\ref{SS:twopoint} we carry out concrete calculations for the
two point correlation function in the $SU(1,1)$ ensemble.
They demonstrate a quadratic repulsion at small distances
and fast decay at infinity. In the limit when the main parameter
$L$ of the $SU(1,1)$ polynomial goes to infinity we recover
the $W_1$ correlation function of Hannay.
There is an appendix at the end of the paper, in which
we consider correlations between outer zeros $z_j$,
$|z_j|>1$, of the $SU(1,1)$ random polynomial. We show that 
in the limit $N\to\infty$, the outer zeros
 are independent of the inner
zeros $|z_j|<1$, and after changing the variable $z$ to $1/z$, the
correlations between outer zeros coincide
with the correlations between zeros in the $SU(1,1)$ ensemble 
with $L=1$.

\section
{Basic Statistical Properties}\label{S:basic}

\subsection{Scaled Density Function}

Consider the following random polynomial, associated with the pseudoeuclidean
group $SU(1,1)$: 
\begin{equation}\label{psi}
\psi(z) = \sum_{m=0}^{N}{\sqrt{C_{m+L-1}^{m}}\;a_{m}z^{m}}\mbox{,}
\qquad C_{m+L-1}^{m}\equiv\frac{(m+L-1)!}{(L-1)!m!}\mbox{,}
\end{equation}
where $L$ is a fixed positive integer, a parameter of the problem, 
$C_{m+L-1}^{m}$ are Newton's binomial coefficients, and $a_m$ are independent standard complex Gaussian random variables, so that
\begin{equation} \label{gauss}
Ea_m=0\mbox{,}\quad Ea_m\overline{a_n}=\delta_{mn}\mbox{,}\quad Ea_ma_n=0\mbox{.}
\end{equation}
When $L=1$, (\ref{psi}) reduces to the classical form
\begin{equation}\label{psi1}
\psi(z) = \sum_{m=0}^{N}a_{m}z^{m}\mbox{.}
\end{equation}
We wish to investigate the density of zeros $p_N(z)$ for the $SU(1,1)$ polynomial (\ref{psi}) as $N \rightarrow \infty$. The density $p_N(z)$ is determined by the condition that for any test function $\phi(z)$, which is infinitely differentiable and compactly supported, 
\begin{equation}
E\left(\sum_{j=1}^N \phi(z_j)\right)=\int_{\C} p_N(z)\phi(z)\,dz\mbox{,}
\qquad dz\equiv dxdy \mbox{,}
\end{equation}
where $z_j$ are zeros of the random polynomial $\psi(z)$. Since the total number of zeros is equal to N,
we have that
\begin{equation}\label{intN}
\int_{\C} p_N(z)\,dz =N\mbox{.}
\end{equation}
Observe that the polynomial
\begin{equation}\label{psie}
\psi(e^{i\theta}z) = \sum_{m=0}^{N}{\sqrt{C_{m+L-1}^{m}}\;a_{m}e^{im\theta}z^{m}}
\end{equation}
has the same probability distribution as $\psi(z)$, hence
\begin{equation}\label{inv}
p_N(re^{i\theta})=p_N(r)\mbox{.}
\end{equation}
The general formula for $p_N(z)$ is given by the Poincar\'e-Lelong type expression 
\begin{equation}\label{pN}
p_N(z) = \frac{1}{\pi}\frac{\partial^2}  {\partial z \partial \bar{z}} [\ln E(\psi(z) \overline{\psi(z)})] 
\end{equation}
(see \cite{BSZ1}, \cite{L}).
From (\ref{psi}),
\begin{equation}\label{Epsi}
E(\psi(z) \overline{\psi(z')})= 
\sum_{m=0}^{N}C_{m+L-1}^{m}(z\overline{ z'})^m 
= \frac{1}{(L-1)!} \sum_{m=0}^{N}{(m+1)\cdot \cdot \cdot (m+L-1)(z\overline{z'})^m} \mbox{.}
\end{equation}
In particular,
\begin{equation}\label{Epsi1}
E(\psi(z) \overline{\psi(z)})={\cal F}_{N, L}(x) 
\equiv \sum_{m=0}^{N}C_{m+L-1}^{m}x^m 
= \frac{1}{(L-1)!} \sum_{m=0}^{N}{(m+1)\cdot \cdot \cdot (m+L-1)x^m} \mbox{.}
\end{equation}
where here and in what follows we use the notation
\begin{equation}
x=z\bar z=|z|^2\mbox{ .}
\end{equation}
It is obvious that
\begin{equation}
{\cal F}_{N, L}(x)>0,\qquad\forall x\ge 0 \mbox{ .}
\end{equation}
Substituting (\ref{Epsi1}) into (\ref{pN}) gives that
\begin{equation}\label{pNF}
p_N(z) = \frac{1}{\pi}\left[x\left(\frac{{\cal F}_{N, L}'(x)}  {{\cal F}_{N, L}(x)}\right)' + \frac{{\cal F}_{N, L}'(x)} {{\cal F}_{N, L}(x)}\right] \mbox{ .}
\end{equation}
We further define
\begin{equation}\label{GNL}
{\cal G}_{N,L}(x) = \sum_{m=0}^{N}{x^{m+L-1}}
     = x^{L-1}\frac{x^{N+1} - 1} {x - 1} \mbox{ .}
\end{equation}
Then from (\ref{Epsi1}),
\begin{equation}\label{FNLFN}
{\cal F}_{N, L}(x) = \frac{1}{(L-1)!} \frac{d^{L-1}{\cal G}_{N,L}(x)} {dx^{L-1}} \mbox{ .}
\end{equation}
Numerical simulations show that most of the roots of $SU(1,1)$ 
polynomials (\ref{psi}) are concentrated in a small annulus (of the width of the order of $1/N$) near 
the unit circle. This property is illustrated in Fig.~1, which contains a point-plot of the zeros of 150 $SU(1,1)$ polynomials of degree $N = 200$, with $L = 30$.  

\vskip.2in
\begin{figure}[ht]\label{scdens0}\centering
\epsfig{file=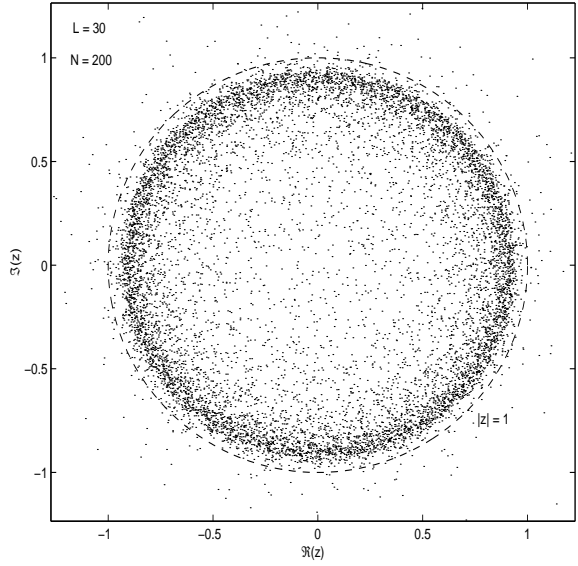, height=3in, width = 3in}
\caption{Zeros of $SU(1,1)$ Random Polynomials}
\end{figure}

To get the asymptotics of the density $p_N(z)$ near the 
unit circle, we introduce a scaling of the variable $x$ in the form 
\begin{equation}\label{xs}
x = 1 + \frac{s}{N}\mbox{ .}
\end{equation}
Then,
\begin{equation}\label{xsd}
\frac{d}{dx}=N\frac{d}{ds} \mbox{ .}
\end{equation}
In subsequent calculations, we will assume that 
\begin{equation}
-A\le s\le A\mbox{ ,}
\end{equation}
for some arbitrary fixed $A>0$. The notation $R(s)=O(N^{-1})$ used below means
that there exists some $C(A)>0$ so that $|R(s)|\le C(A)N^{-1}$ for all $s\in[-A,A]$.
The main result of this section is the following theorem.

\begin{theo}\label{density} As $N\to\infty$,
\begin{equation} \label{pNas0}
N^{-2}p_N\left(\left(1+\frac{s}{N}\right)^{1/2}e^{i\theta}\right)
=\frac{1}{\pi}\left[\frac{g^{(L)}(s)}{g^{(L-1)}(s)}\right]'+O(N^{-1})\mbox{ .}
\end{equation}
where
\begin{equation} \label{gj0}
g^{(j)}(s)=\frac{d^j}{ds^j}\left(\frac{e^s-1}{s}\right)\mbox{ .}
\end{equation}
\end{theo}

{\it Remark.} Observe that one $N$ in the normalization of $p_N(z)$ on the left
is due to the rescaling (\ref{xs}) and one is due to the integral condition (\ref{intN}).

{\it Proof.}
By substituting (\ref{xs}) into (\ref{GNL}) we obtain that as $N\to\infty$,
\begin{equation}\label{FNAS}
{\cal G}_{N,L}(x) = \left(1 + \frac{s}{N}\right)^{L-1} \: \frac{(1 + \frac{s}{N})^{N+1} - 1} {\frac{s}{N}}
     = N \frac{e^s - 1}{s}\left(1+O(N^{-1})\right) = Ng(s)\left(1+O(N^{-1})\right)\mbox{ ,}
\end{equation}
where we define
\begin{equation}\label{g}
g(s) = \frac{e^s - 1}{s}=\sum_{j=0}^\infty \frac{s^j}{(j+1)!} \mbox{ .}
\end{equation}
Observe that for $j=0,1,2,\dots$,
\begin{equation}\label{gj}
g^{(j)}(s)\equiv \frac{d^jg(s)}{ds^j}>0\mbox{ ,}\qquad -\infty< s<\infty  \mbox{ .}
\end{equation}
It is obvious from (\ref{g}) for $s\ge 0$. For negative $s$, use the identity
\begin{equation} \label{gj-}
g^{(j)}(-s)=\frac{j!}{s^{j+1}}\left[1-e^{-s}\left(1+\frac{s}{1!}+\dots+\frac{s^j}{j!}\right)\right] \mbox{ .}
\end{equation}
Formula (\ref{FNAS}) holds obviously in a small complex neighborhood of the segment
$-A\le s\le A$. Therefore, we can differentiate it in $s$, so that
\begin{equation}\label{FNASD}
\frac{d^j{\cal G}_{N,L}(x)}{dx^j}
 =  N^{j+1}g^{(j)}(s)\left(1+O(N^{-1})\right)\mbox{ ,}
\end{equation}  
Thus, from (\ref{FNLFN}), we obtain that
\begin{equation}
{\cal F}_{N, L}(x)= \frac{N^Lg^{(L-1)}(s)}{(L-1)!}\left(1+O(N^{-1})\right)\mbox{,} \qquad 
g^{(L-1)}(s)\equiv\frac{d^{L-1}g(s)}{ds^{L-1}}\mbox{ ,}
\end{equation}
and 
\begin{equation}
{\cal F}'_{N, L}(x)= \frac{N^{L+1}g^{(L)}(s)}{(L-1)!}\left(1+O(N^{-1})\right)\mbox{ ,}
\end{equation}
hence
\begin{equation}
\frac{{\cal F}'_{N, L}(x)}{{\cal F}_{N, L}(x)}= \frac{Ng^{(L)}(s)}{g^{(L-1)}(s)}\left(1+O(N^{-1})\right)\mbox{ .}
\end{equation}
Let us go back to formula (\ref{pNF}). In this formula, $x=1+O(N^{-1})$, and the first
term in the brackets has an extra derivative so this is the leading term,
and we can neglect the second term. This gives that
\begin{equation} \label{pNas1}
N^{-2}p_N(z)=\frac{1}{\pi}\left[\frac{g^{(L)}(s)}{g^{(L-1)}(s)}\right]'+O(N^{-1})\mbox{,}
\qquad z\bar z=1+\frac{s}{N}\mbox{ ,}
\end{equation}
which was stated. Theorem \ref{density} is proved.

Equation (\ref{pNas0}) can also be expressed in the following form:
\begin{equation}\label{pNas2}
N^{-2}p_N(z)=\frac{1}{\pi}\left[\log g^{(L-1)}(s)\right]''+O(N^{-1})\mbox{.}
\end{equation}
From (\ref{pNas0}),
\begin{equation} \label{p}
\lim_{N\to\infty}N^{-2}p_N\left(\left(1+\frac{s}{2N}\right)e^{i\theta}\right)
=p(s)\equiv\frac{1}{\pi}\left[\frac{g^{(L)}(s)}{g^{(L-1)}(s)}\right]'
=\frac{1}{\pi}\left[\log g^{(L-1)}(s)\right]''\mbox{.}
\end{equation}

Consider now the distribution function of zeros. Define
\begin{equation}\label{PN}
P_N(x)=N^{-1}E\#\{j: |z_j|^2\le x\}\mbox{,}
\end{equation}
which gives the expected value of the fraction of zeros in the disk of radius $\sqrt x$.
Then
\begin{equation}\label{PND}
P'_N(x)=N^{-1}\pi p_N(z)\mbox{,}\qquad x=|z|^2\mbox{.}
\end{equation}
Hence (\ref{pNas0}) implies that
\begin{equation} \label{PNas1}
P_N\left(1+\frac{s}{N}\right)=\frac{g^{(L)}(s)}{g^{(L-1)}(s)}+O(N^{-1})\mbox{.}
\end{equation}
As follows from (\ref{gj-}), the limiting distribution function,
\begin{equation} \label{P}
\lim_{N\to\infty}P_N\left(1+\frac{s}{N}\right)=P(s)\equiv\frac{g^{(L)}(s)}{g^{(L-1)}(s)}\mbox{,}
\end{equation}
has the following asymptotics at $-\infty$:
\begin{equation} \label{P-}
P(s)=-\frac{L}{s}+O\left((-s)^{L-1}e^{s}\right)\mbox{,}\qquad s\to-\infty\mbox{.}
\end{equation}
At $+\infty$ we use the formula
\begin{equation} \label{gj+}
g^{(j)}(s)=\frac{e^s}{s}\left[1-\frac{j}{s}+\frac{j(j-1)}{s^2}-\dots+\frac{(-1)^jj!}{s^j}\right]
+\frac{(-1)^{j+1}j!}{s^{j+1}}\mbox{,}
\end{equation}
which gives that
\begin{equation} \label{P+}
P(s)=\frac{1}{s}-\frac{L-1}{s^2}+O\left(s^{-3}\right)\mbox{,}\qquad s\to \infty\mbox{.}
\end{equation}
For $L=1$, (\ref{P}) reduces to the well-known result (see \cite{BS})
\begin{equation} \label{PNas11}
P(s)=\frac{se^s-e^s+1}{s(e^s-1)}\mbox{.}
\end{equation}
For $L=1$, the limiting distribution is symmetric, so that
\begin{equation} \label{symm}
P(1-s)=1-P(s)
\end{equation}
(it is related to the symmetry of the zeros of $\psi(z)$ in (\ref{psi1}) with respect to
the transformation $z\to1/z$).  
Since
\begin{equation} \label{gseries}
g(s)=1+\frac{s}{2!}+\frac{s^2}{3!}+\frac{s^3}{4!}+\dots\mbox{,}
\end{equation}
we have that
\begin{equation} \label{gjseries}
g^{(j)}(s)=\frac{1}{j+1}+\frac{s}{1!(j+2)}+\frac{s^2}{2!(j+3)}
+\frac{s^3}{3!(j+4)}+\dots\mbox{.}
\end{equation}
Thus,
\begin{equation} \label{PNas2}
P_N\left(1+\frac{s}{N}\right)=\frac{\displaystyle
\frac{1}{L+1}+\frac{s}{1!(L+2)}+\frac{s^2}{2!(L+3)}+\dots}
{\displaystyle
\frac{1}{L}+\frac{s}{1!(L+1)}+\frac{s^2}{2!(L+2)}+\dots}+O(N^{-1})\mbox{.}
\end{equation}
In particular,
\begin{equation} \label{PNas3}
P_N(1)=\frac{L}{L+1}+O(N^{-1})\mbox{,}
\end{equation}
which means that the expected value of the fraction of zeros inside 
the unit disk is asymptotically
equal to $L/(L+1)$.

In our numerical simulations we generated a large number of $SU(1,1)$ 
random polynomials of degree $N$, calculated the zeros using standard 
techniques, and counted the number of zeros in annuli of fixed width, 
which are concentrically spread around the origin, per one polynomial. 
The scaled density was obtained by dividing each of these numbers 
by the area of the corresponding annulus, as well as by adjusting 
it with respect to $N$ (according to the previously established result). 
Figure~2 shows the results of this procedure for $L = 4$ and $N = 150$, 
in comparison with the theoretical value as given by Eq. (\ref{pNas0}).
         
\begin{figure}[ht]\label{scdens}\centering
\epsfig{file=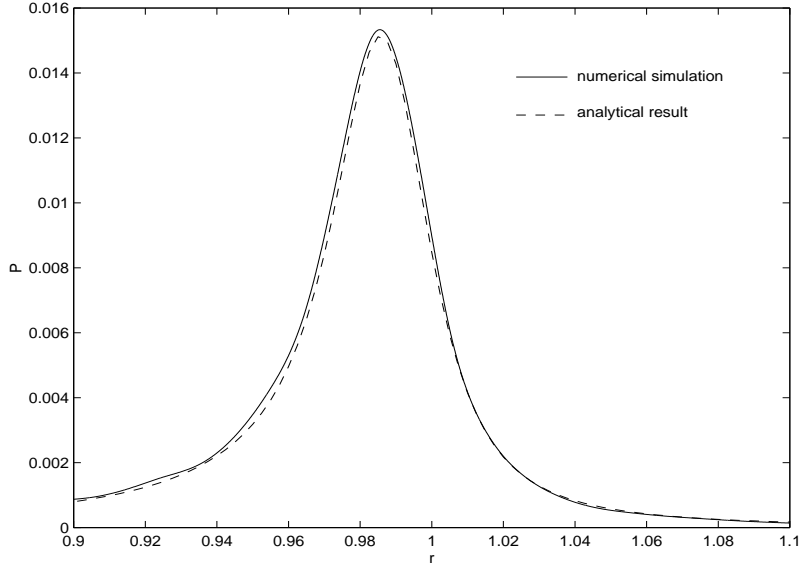, height=3in, width = 4.2in}
\caption{The Scaled Density Function - Theoretical Limit 
and Computer Simulation}
\end{figure}

\subsection{$SU(1,1)$ Ensemble of Random Analytic Functions} \label{SS:SU11}

In this section we consider the $SU(1,1)$ random analytic function, 
\begin{equation}\label{psi2}
\psi(z) = \sum_{m=0}^{\infty}{\sqrt{C_{m+L-1}^{m}}a_{m}z^{m}}\mbox{,}
\end{equation}
which is obtained from (\ref{psi}) by setting $N=\infty$. Here again
$C_{m+L-1}^{m}$ are Newton's binomial coefficients, and $a_m$ are independent standard complex Gaussian random variables. 

\begin{prop} \label{covprop}
 Random series (\ref{psi2}) converges almost surely for all $|z|<1$.
For all mutually distinct $z_1,\dots,z_n,$ in the
disk $|z|<1$ the covariance matrix
\begin{equation}\label{covA}
A_n=\left( E\left(\psi(z_p)\overline{\psi(z_{p'})}\right)\right)_{p,p'=1,\dots,n}
\end{equation}
is positive definite.
\end{prop}

{\it Proof. Almost sure convergence.} Observe that $(m+j)/m\le j+1$ hence
\begin{equation}\label{prop1}
\frac{(m+1)\dots (m+L-1)}{m^{L-1}}\le L!
\end{equation}
and
\begin{equation}\label{prop2}
C_{m+L-1}^m=\frac{(m+L-1)\dots (m+1)}{(L-1)!}\le Lm^{L-1} \mbox{.}
\end{equation}
Consider the set
\begin{equation}\label{prop3}
\Lambda_N=\{a=(a_0,a_1,\dots)\,:\; |a_m|\le m\quad \forall\, m\ge N\}\mbox{.}
\end{equation}
Then for all $a\in\Lambda_N$ series (\ref{psi2}) converges and
\begin{equation}\label{prop4}
\lim_{N\to\infty}\Prob \,\Lambda_N=1\mbox{,}
\end{equation}
which implies the almost sure convergence.

{\it Positive definiteness.} Consider the quadratic form
\begin{equation}\label{prop5}
A_n(\mu)=\sum_{p,p'=1}^n E\left(\psi(z_p)\overline{\psi(z_{p'})}\right) \mu_p
\overline{\mu_{p'}}=E\left|\sum_{p=1}^n \mu_p\psi(z_p)\right|^2\mbox{.}
\end{equation}
Assume that $A_n$ is not positive definite. Then, for some nonzero
vector $\mu=(\mu_1,\dots,\mu_n)$,
\begin{equation}\label{prop6}
\sum_{p=1}^n \mu_p\psi(z_p)=0
\end{equation}
for almost all $\psi$. Hence, (\ref{prop6}) holds for almost $\psi$
with coefficients vector $a\in\Lambda_N$. Assume that $N>n$.
There exists a polynomial
\begin{equation}\label{prop7}
\psi_0(z)=\sum_{m=0}^{n-1} \sqrt{ C_{m+L-1}^m}\,a_m^0\,z^m\mbox{,}
\end{equation}
such that
\begin{equation}\label{prop8}
\psi_0(z_p)=\overline{\mu_p}\mbox{,}
\end{equation}
so that
\begin{equation}\label{prop9}
\sum_{p=1}^n \mu_p\psi_0(z_p)=\sum_{p=1}^n |\mu_p|^2\not= 0\mbox{.}
\end{equation}
Let $a^0=(a_0^0,a_1^0,\dots,a_{n-1}^0,0,0,\dots)$. From (\ref{prop3}),
it is obvious that if $a\in\Lambda_N$ then $(a+a^0)\in\Lambda_N$.
The shift $a\to a+a^0$ is a measurable one-to-one transformation
in $\Lambda_N$, hence for almost all vectors $a\in\Lambda_N$ we have
that (\ref{prop6}) holds along with
\begin{equation}\label{prop10}
\sum_{p=1}^n \mu_p[\psi(z_p)+\psi_0(z_p)]=0\mbox{.}
\end{equation}
But this implies that
\begin{equation}\label{prop11}
\sum_{p=1}^n \mu_p\psi_0(z_p)= 0\mbox{,}
\end{equation}
which is in contradiction with (\ref{prop9}). This contradiction proves
the positive definiteness of $A_n$. Proposition \ref{covprop} is proved.

{\it Remark.} The above proof gives also the positive definiteness
of the covariance matrix $A_n$ corresponding to the random polynomial (\ref{psi}),
provided $N>n$.

In this section, we will be interested in the density function 
$\rho(z)$ of the distribution of zeros of (\ref{psi2})
in the disk $|z| < 1$.  We have from (\ref{psi1}) that
\begin{equation} \label{Epsipsi}
E(\psi(z) \overline{\psi(z')}) = \sum_{m=0}^{\infty}{{C_{m+L-1}^{m}}(z \overline{z'})^m}
 =\frac{1}{(1-z\overline{z'})^L}\mbox{ .}
\end{equation} 
In particular,
\begin{equation} \label{Epsi2}
E(\psi(z) \overline{\psi(z)}) = f(x)\equiv \frac{1}{(1-x)^L} 
\mbox{,}
\qquad x\equiv z\bar z=|z|^2 \mbox{ .}
\end{equation}
In terms of $f$ the density function $\rho(z)$ of the zeros of
(\ref{psi2}) is given by: 
\begin{equation}\label{rho}
\rho(z) = \frac{1}{\pi}\left[x\left(\frac{f'(x)}{f(x)}\right)' +
\frac{f'(x)}{f(x)}\right]\mbox{ .} 
\end{equation}
Substituting (\ref{Epsi2}) into this formula gives that
\begin{equation}
\rho(z) = \frac{1}{\pi}\left[x\left(\frac{L}{1-x}\right)' +
\frac{L}{1-x}\right] 
=\frac{L}{\pi(1-x)^2}\mbox{ ,}
\end{equation}
or going back to the complex variable $z$, we obtain the result of
Leboeuf \cite{L}, 
\begin{equation}\label{rhoz}
\rho(z) = \frac{L}{\pi\left(1 - |z|^2\right)^2} \mbox{ .}
\end{equation}
Numerical simulations were similar to those of the previous
section. The density was simply obtained by dividing the number of
counted zeros by the area of the corresponding annulus. Figure~3 shows
the results of this procedure for the degree $N$ equal to 50, 75, 100,
and 150, in comparison with the theoretical value as given by
Eq. (\ref{rhoz}).  

\begin{figure}[ht]\label{unscdens}\centering
\epsfig{file=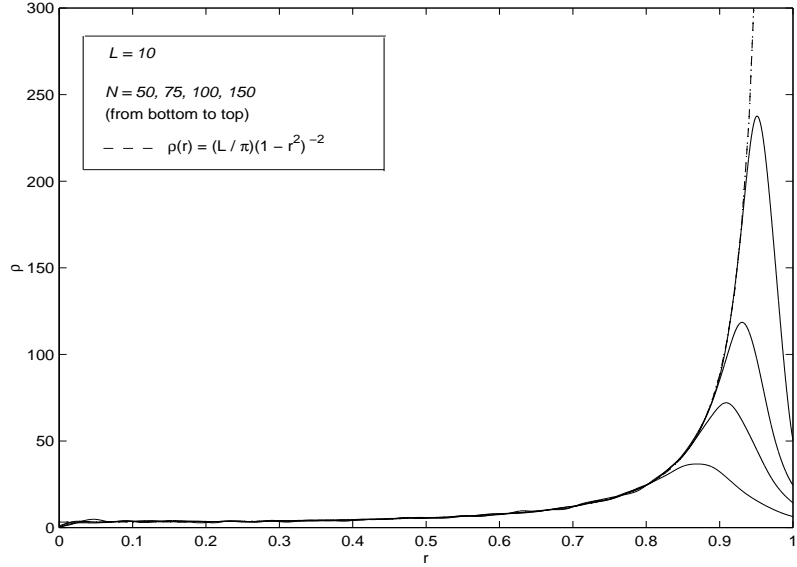, height=3in, width = 4.2in}
\caption{The Unscaled Density Function - Theoretical Limit and
  Computer Simulation} 
\end{figure}

\section {Correlations between Zeros} \label{S:corr}

\subsection{The $SU(1,1)$ Symmetry}

We have the following important property:
\begin{theo} \label{invariance} The distribution of zeros of the random
  analytic function (\ref{psi2}) 
is invariant with respect to the action of the group $SU(1,1)$,
\begin{equation}\label{su}
z\to \frac {az+b}{cz+d},\qquad
\left(\begin{array}{cc}
a & b \\ c & d
\end{array}\right)
\in SU(1,1) \mbox{ .}
\end{equation}
\end{theo}

{\it Remark.} This implies that all joint distribution functions of
zeros of $\psi(z)$ are $SU(1,1)$ invariant. 

{\it Proof.} Consider the following homogeneous analytic function of
two variables: 
\begin{equation}\label{Psi2}
\Psi(z_0,z_1) =
z_0^{-L}\sum_{m=0}^{\infty}{\sqrt{C_{m+L-1}^{m}}a_{m}
\left(\frac{z_1}{z_0}\right)^{m}}\mbox{,}  
\qquad |z_1|<|z_0| \mbox{.}
\end{equation}
Then
\begin{equation}\label{psiPsi}
\Psi(1,z_1) = \psi(z_1)\mbox{.}
\end{equation}
Let us find the covariance function of $\Psi(z_0,z_1)$:
\begin{eqnarray}\label{covPsi}
E\left(\Psi(z_0,z_1)\overline{\Psi(z'_0,z'_1)}\right)
 &=&
 \left(z_0\overline{z'_0}\right)^{-L}
\sum_{m=0}^{\infty}{{C_{m+L-1}^{m}}\left(\frac{z_1
 \overline{z_1'}}{z_0 \overline{z_0'}}\right)^m}\\ 
&=&\left(z_0\overline{z'_0}\right)^{-L}
\left(1-\frac{z_1 \overline{z_1'}}{z_0 \overline{z_0'}}\right)^{-L}
=\left(z_0 \overline{z_0'}-z_1 \overline{z_1'}\right)^{-L}
\mbox{.}
\end{eqnarray}
The action of a matrix $A\in SU(1,1)$ preserves the (1,1) indefinite
scalar product $z_0 \overline{z_0'}-z_1 \overline{z_1'}$.
Thus, the covariance function of the Gaussian random analytic function
$\Psi(z_0,z_1)$ 
is $SU(1,1)$ invariant. In addition,
\begin{equation}\label{EPsi}
E\Psi(z_0,z_1) = 0\mbox{,}\qquad E\Psi(z_0,z_1)\Psi(z'_0,z'_1)=0\mbox{.}
\end{equation}
For the Gaussian random function, the first two moments determine it uniquely.
This implies that the distribution of zeros of $\Psi(z_0,z_1)$ is
$SU(1,1)$ invariant. Restricting this to $z_0=1$ we get that the
distribution of zeros of $\psi(z)$ is $SU(1,1)$ invariant as well.
Theorem \ref{invariance} is proved.

Recall that a general formula for a matrix $A\in SU(1,1)$ is the following:
\begin{equation}\label{A}
A=\left(\begin{array}{cc}
a & b \\ \overline b & \overline a
\end{array}\right)
\mbox{,}\qquad |b|^2-|a|^2=1\mbox{,}
\end{equation}
so that it depends on three real parameters. An $SU(1,1)$-invariant
metric is
\begin{equation}\label{metric}
ds^2=\frac{4(dx^2+dy^2)}{(1-x^2-y^2)^2}
\mbox{,}
\end{equation}
and the corresponding distance $\tau(z_1,z_2)$ on 
the disk $\{|z|<1\}$ is determined by the equation
\begin{equation}\label{distance}
\tanh(\tau/2)=\frac{|z_1-z_2|}{|1-z_1\overline{z_2}|}
\mbox{.}
\end{equation}
The corresponding $SU(1,1)$-invariant volume element is
\begin{equation}\label{volume}
\frac{4dxdy}{(1-x^2-y^2)^2}
\mbox{.}
\end{equation}
Theorem \ref{invariance}
implies the $SU(1,1)$ invariance for the normalized correlation functions.\\

\subsection{Correlation Functions - Preliminaries}

The $n$-point correlation function $K_n(z_1,\dots,z_n)$ is determined
by the condition 
that for any test functions $\phi_1(z),\dots,\phi_n(z)$, which are infinitely 
differentiable and compactly supported in the disk $\{|z|<1\}$, such
that their supports do not intersect,  
\begin{equation}\label{Kn}
E\prod_{p=1}^n\left(\sum_{j=1}^{\infty} \phi_p(\zeta_j)\right)
=\int_{\C^n} K_n(z_1,\dots,z_n)\prod_{p=1}^n [\phi(z_p)\,dz_p]\mbox{,}
\end{equation}
where $\zeta_j$ are zeros of the $SU(1,1)$ random analytical function $\psi(z)$
(we denote them by $\zeta_j$ to distinguish them from the variables $z_j$).
The sum over $j$ is, in fact, finite because $\phi_p(z)$ has a compact support.
The general formula for $K_n(z_1,\dots,z_n)$ is given by the Kac-Rice
expression 
\begin{equation}\label{KacRice}
K_n(z)=\int d\xi\, D_n(0,\xi;z)\prod_{p=1}^n (\xi_p\xi_p^{*})
\mbox{,}\qquad z=(z_1,\dots, z_n)\mbox{,}\qquad \xi=(\xi_1,\dots,\xi_n)
\mbox{,}
\end{equation}
where $D_n(x,\xi;z)$, $x,\xi\in C^n$, is the distribution density of
the two random 
vectors
\begin{equation}\label{rv}
X=\left(\psi(z_1),\dots,\psi(z_n)\right)\mbox{,}\qquad 
\Xi=\left(\psi'(z_1),\dots,\psi'(z_n)\right)\mbox{.}
\end{equation}
Formula (\ref{KacRice}) is derived in \cite{BSZ2} in a much more general
situation of sections of powers of a line bundle over a complex manifold,
as a generalization of
the original formula by Kac and Rice \cite{Ka}, \cite{Ri} (see also \cite
{H} and \cite{BD}).

For the Gaussian random functions,  formula (\ref{KacRice}) can be specialized
as follows (see \cite{H}, \cite{BD}, and \cite{BSZ2}):
\begin{equation}\label{Knz}
K_n(z)=\frac{1}{\pi^n\det
  A_n}\left\langle\prod_{p=1}^n(\xi_p\xi_p^{*})\right\rangle_{\Lambda_n} 
\mbox{ ,}
\end{equation}
where   
\begin{equation}\label{An}
A_n=\left( E\,\psi(z_p)\overline{\psi(z_{p'})}\right)_{p,p'=1,\dots,n}\mbox{,}
\end{equation}
and $\langle\cdot\rangle_{\Lambda_n}$ stands for averaging with respect to
Gaussian complex random variables $\xi_1,\dots,\xi_n$, such that 
\begin{equation}\label{xixi}
\left(E\,\xi_p\overline{\xi_{p'}}\right)_{p,p'=1,\dots,n}
=\Lambda_n\mbox{,}\qquad 
E\,\xi_p=0\mbox{,}\qquad E\,\xi_p\xi_{p'}=0\mbox{,}\qquad
p,p'=1,\dots,n\mbox{,} 
\end{equation}
where
\begin{equation}\label{Lam}
\Lambda_n=C_n-B^*_nA_n^{-1}B_n
\end{equation}
and
\begin{equation}\label{BnCn}
B_n=\left(
E\,\psi(z_p)\overline{\psi'(z_{p'})}\right)_{p,p'=1,\dots,n}\mbox{,}\qquad 
C_n=\left(
E\,\psi'(z_p)\overline{\psi'(z_{p'})}\right)_{p,p'=1,\dots,n}\mbox{.} 
\end{equation}
For $n=1$, formula (\ref{Knz}) reduces to
\begin{equation}\label{K1z}
K_1(z)=\frac{1}{\pi}\,
\frac{A_n(z)C_n(z)-B_n(z)\overline{B_n(z)}}{A_n^2(z)}\mbox{,} 
\end{equation}
where
\begin{equation}\label{ABC}
A_n(z)=E\,\psi(z)\overline{\psi(z)}
\mbox{,}\qquad B_n(z)=\frac{\partial A_n(z)}{\partial z}
\mbox{,}\qquad C_n(z)=\frac{\partial^2 A_n(z)}{\partial z
  \partial\overline{  z}} 
\mbox{.}
\end{equation}
The 1-point correlation function $K_1(z)$ is nothing else than the
density function  
of the distribution of zeros and
formula (\ref{K1z}) is equivalent to the Poincar\'e-Lelong type
formula (\ref{pN}). 

The normalized correlation function $k_n(z_1,\dots,z_n)$ is defined as
\begin{equation}\label{kn}
k_n(z_1,\dots,z_n)=\frac{K_n(z_1,\dots,z_n)}{K_1(z_1)\dots K_1(z_n)}\mbox{.}
\end{equation}
It satisfies the following theorem.

\begin{theo} \label{kninv} The function $k_n(z_1,\dots,z_n)$
is invariant with respect to the action of the group $SU(1,1)$,
\begin{equation}\label{kinv}
k_n(Az_1,\dots,Az_n)=k_n(z_1,\dots,z_n) \mbox{ ,}\qquad \forall \,A\in
SU(1,1)\mbox{,} 
\end{equation}
where
\begin{equation}\label{Az}
Az=\frac{az+b}{cz+d}\mbox{.}
\end{equation}
\end{theo}

{\it Remark.} This implies that $k_n(z_1,\dots,z_n)$ is a function of 
pairwise distances $\tau(z_p,z_q)$ defined by (\ref{metric}).

{\it Proof.} By Theorem \ref{invariance}, the distribution
$K_n(z_1,\dots,z_n)dz_1\dots dz_n$ 
is $SU(1,1)$ invariant. Also, the distribution $K_1(z_1)\dots
K_1(z_n)dz_1\dots dz_n$ 
is $SU(1,1)$ invariant. Hence their quotient (the Radon~-~Nikodim derivative),
\begin{equation}\label{RadNik}
\frac{K_n(z_1,\dots,z_n)dz_1\dots dz_n}{K_1(z_1)\dots K_1(z_n)dz_1\dots dz_n}=
k_n(z_1,\dots,z_n)
\end{equation}
is $SU(1,1)$ invariant, which was stated. Theorem \ref{kninv} is proved.

The normalized correlation function can be expressed as a supersymmetric (Berezin)
integral,
\begin{equation}\label{super}
k_n(z_1,\dots,z_n)=\frac 1{\det A_n}\int\frac 1{\det [I+\Lambda_n(z)\Omega]}\,d\eta
\mbox{,}
\end{equation}
where $\Omega$ is $n\times n$ matrix,
\begin{equation}\label{Omega}
\Omega=\left(\delta_{pp'}\eta_{p'}\bar\eta_{p}\right)_{p,p'=1,\dots,n}
\mbox{,}
\end{equation}
and the $\eta_p,\;\bar\eta_p$ are anti-commuting (fermionic) variables,
with $d\eta=\prod_p d\bar\eta_p d\eta_p$. The integral in (\ref{super})
is a Berezin integral, which is evaluated by simply taking
the coefficient of the top degree form of the integrand
$\frac 1{\det [I+\Lambda_n(z)\Omega]}$. For a derivation of formula (\ref{super}), see
\cite{BSZ4}.

\subsection{The Two-Point Correlation Function}\label{SS:twopoint}

From (\ref{psi2}), (\ref{rhoz}), (\ref{An}), and (\ref{BnCn}), we can
directly obtain the following expressions for $A_n$, $B_n$, and $C_n$
in the case of a two-point correlation function $k_{2}(z_{1}, z_{2})$: 
\begin{eqnarray}
A_{2}& = &\left(\begin{array}{cc}
(1 - |z_{1}|^{2})^{-L} & (1 - z_{1} \overline{z_{2}})^{-L} \\ \noalign{\medskip}
(1 - \overline{z_{1}} z_{2})^{-L} & (1 - |z_{2}|^{2})^{-L}
\end{array}\right) \nonumber \\ \noalign{\medskip}
B_{2}& = &\left(\begin{array}{cc}
Lz_{1}(1 - |z_{1}|^{2})^{-L-1} & Lz_{1}(1 - z_{1} \overline{z_{2}})^{-L-1}
\\ \noalign{\medskip} 
Lz_{2}(1 - \overline{z_{1}} z_{2})^{-L-1} & Lz_{2}(1 - |z_{2}|^{2})^{-L-1}
\end{array}\right) \label{ABC_2}\\ \noalign{\medskip}
C_{2}& = &\left(\begin{array}{cc}
L(1 + L|z_{1}|^{2})(1 - |z_{1}|^{2})^{-L-2} & L(1 + Lz_{1}
\overline{z_{2}})(1 - z_{1} \overline{z_{2}})^{-L-2} \\ \noalign{\medskip} 
L(1 + L\overline{z_{1}} z_{2})(1 - \overline{z_{1}} z_{2})^{-L-2} & L(1 +
L|z_{2}|^{2})(1 - |z_{2}|^{2})^{-L-2} 
\end{array}\right) \nonumber \mbox{ .}
\end{eqnarray}\\
By Theorem \ref{kninv}, $k_2(z_1,z_2)$ is $SU(1,1)$ invariant,
that is
\begin{equation}\label{k2inv}
k_2(Az_1,Az_2)=k_2(z_1,z_2)\mbox{,}\qquad\forall\, 
A\in SU(1,1)\mbox{.}
\end{equation}
The action of $SU(1,1)$ is transitive, hence we can move $z_1$
to the origin, and by rotation, we can then move $z_2$ to the positive
half-axis. Therefore, we will assume that $z_1=0$ and $z_2=r>0$.
 In that case, equations (\ref{ABC_2}) become 
\begin{eqnarray}
A_{2}& = &\left(\begin{array}{cc}
1 & 1 \\ \noalign{\medskip}
1 & (1 - r^{2})^{-L}
\end{array}\right) \nonumber \\ \noalign{\medskip}
B_{2}& = &\left(\begin{array}{cc}
0 & 0 \\ \noalign{\medskip}
Lr & Lr(1 - r^{2})^{-L-1}
\end{array}\right) \label{ABC_2r}\\ \noalign{\medskip}
C_{2}& = &\left(\begin{array}{cc}
L & L \\ \noalign{\medskip}
L & L(1 + Lr^{2})(1 - r^{2})^{-L-2}
\end{array}\right) \nonumber \mbox{ .}
\end{eqnarray}\\
From (\ref{ABC_2r}) and (\ref{Lam}), we further obtain
\begin{equation}\label{Lam_n_0}
\Lambda_{2} = \left(\begin{array}{cc}
L - L^{2}r^{2}\frac{1}{(1-r^{2})^{-L}-1} & L -
L^{2}r^{2}\frac{(1-r^{2})^{-L-1}}{(1-r^{2})^{-L}-1} \\
\noalign{\medskip} 
L - L^{2}r^{2}\frac{(1-r^{2})^{-L-1}}{(1-r^{2})^{-L}-1} & L(1 +
Lr^{2})(1 - r^{2})^{-L-2} -
L^{2}r^{2}\frac{(1-r^{2})^{-2L-2}}{(1-r^{2})^{-L}-1} 
\end{array}\right) \mbox{ .}
\end{equation}
For the two-point case, formula (\ref{Knz}) reduces to the following
expression: 
\begin{equation}\label{K2z}
K_2(z_1, z_2)=\frac{1}{\pi^2\det A_2}\left(\Lambda_{2, 11}\Lambda_{2,
  22} + \Lambda_{2, 12}\Lambda_{2, 21}\right) 
\mbox{ ,}
\end{equation} 
which, when combined with (\ref{kn}), (\ref{K1z}), and (\ref{rhoz}),
results in the following formula for the normalized two-point
correlation function:
\begin{eqnarray}\label{twopoint}
k_2(z_{1}, z_{2}) &=&\left[ \left(1-{r}^{2}\right)^{3L+2} +
 \left((L^2-2L-2)r^{4} + (4L+4)r^{2} - 1\right)
 \left(1-{r}^{2}\right)^{2L}  \right. \nonumber \\ 
 & & + \left. \left((L+1)^{2}r^{4} - (4L+2)r^{2} - 1\right)
 \left(1-{r}^{2}\right)^{L} + 1 \right] \left/ \left(1-\left
 (1-{r}^{2}\right )^{L}\right )^{3} \right. \mbox{ ,} 
\end{eqnarray}
where according to (\ref{distance}), 
\begin{equation}\label{rtau}
r = \tanh\left(\frac{\tau} {2}\right) =\frac
{|z_1-z_2|}{\left|1-z_1\overline{z_2}\right|}\mbox{.} 
\end{equation}
When $L=1$, formula (\ref{twopoint}) simplifies to
\begin{equation}\label{twop1}
k_2(z_1,z_2)=r^2(2-r^2)\mbox{,}\qquad L=1 
\mbox{.} 
\end{equation}
Plots of $k_2\left(\tanh\left(\frac{\tau} {2}\right)\right)$ for
$L=1,5$ and 50 are shown in Fig.~4.

\begin{figure}[htb]\label{k2plot0}\centering
\epsfig{file=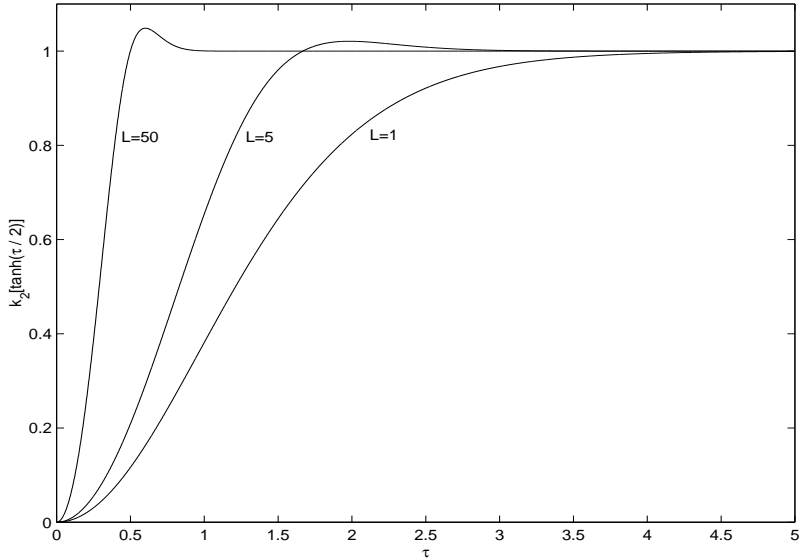, height=3in, width = 4.2in}
\caption{The Two-Point Correlation Function} 
\end{figure}

As it can be seen from the plot, the two-point correlation function
goes to 0 as $\tau \rightarrow 0$ (or, in other words, as $r
\rightarrow 0$). The limiting behavior can be obtained through a
series expansion of (\ref{twopoint}). The following expression was
obtained using Maple$^{\rm TM}$: 
\begin{equation}\label{r->0}
k_{2}(r) = \frac{1}{2}\frac{(L+1)^2}{L}r^2 - \frac{1}{4}\frac{(L+1)^2}{L}r^4 -
           \frac{1}{36}\frac{(L^2-1)^2}{L}r^6 -
           \frac{1}{72}\frac{(L^2-1)^2}{L}r^8 + O(r^{10}) \mbox{ ,} 
\end{equation} 
which shows dominating quadratic behavior in the neighborhood of $r =0$. Thus,
there is a quadratic repulsion between zeros.

We are also interested in the asymptotic behavior of the correlation
function as $L \rightarrow \infty$. For this purpose, we introduce the
scaling 
\begin{equation}\label{uLscale}
r = \frac{u}{\sqrt{L}} \mbox{ .}
\end{equation}
Substituting this into (\ref{twopoint}) and taking the
limit $L\to \infty$, we obtain the following expression:
\begin{equation}\label{explim}
k_{2}(u) = {\frac {{e^{3\,{u}^{2}}}+({u}^{4}-4u^2-1)e^{2\,{u}^{2}}
+(u^4+4u^2-1)e^{{u}^{2}}+1}{\left ({e^{{u}^{2}}}-1\right )^{3}}} \mbox{ ,}
\end{equation}
which can also be written as
\begin{equation}\label{sinhlim}
k_2(u) = \frac {\left ( \sinh^{2}t + t^2
\right )\cosh t-
2t\sinh t}{ \sinh^{3}t } \mbox{ ,}\qquad t=\frac{u^2}2 \mbox{ ,}
\end{equation}
and agrees with the result obtained by Hannay \cite{H}. 
A plot of (\ref{sinhlim}) is shown in Fig.~5.

\begin{figure}[ht]\label{k2plot}\centering
\epsfig{file=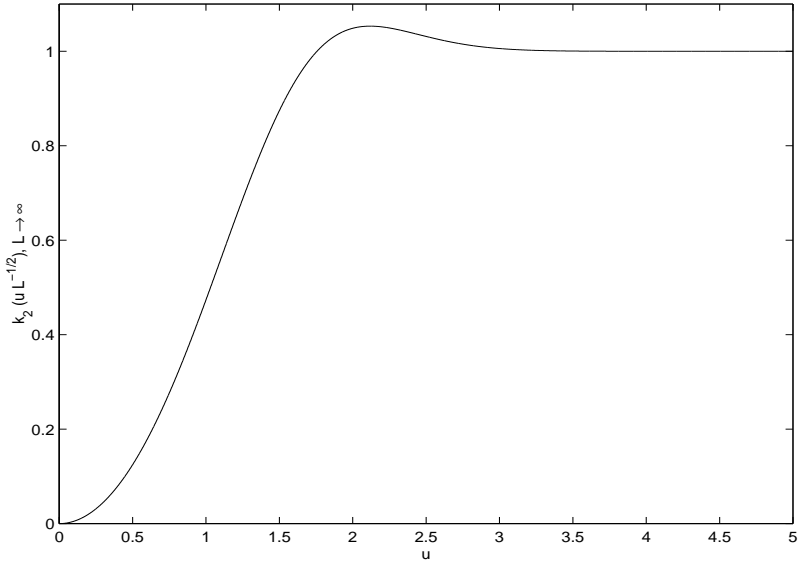, height=3in, width = 4.2in}
\caption{Asymptotics of the Two-Point Correlation Function as $L
  \rightarrow \infty$}  
\end{figure}
 
In fact, under the scaling (\ref{uLscale}), the $SU(1,1)$ random analytic
function converges, as $L\to\infty$, to the $W_1$ random analytic function
(cf. \cite{L}). Indeed,
\begin{equation}\label{infi}
\psi\left(\frac{u}{\sqrt L}\right)
=\sum_{m=0}^{\infty}\sqrt{\frac{L(L+1)\dots (L+m-1)}{L^m m!}}\;a_{m}u^{m}\mbox{.}
\end{equation}
As $L\to\infty$, the expression under the radical approaches $1/m!$, so that
$\psi(u/\sqrt{L})$ approaches 
\begin{equation}\label{infi2}
\psi_0(u)=\sum_{m=0}^{\infty}\sqrt{\frac{1}{m!}}\;a_{m}u^{m}\mbox{,}
\end{equation}
which is the $W_1$ random analytic function.

\appendix
\section { Appendix. Correlations between Inner and Outer Zeros} 
\label{S:outer}

In this appendix, we consider the limit of correlations between zeros of
the random polynomial (\ref{psi}) as $N\to\infty$. In the open disk
$\{ |z|<1\}$, the random polynomial (\ref{psi}) approaches the $SU(1,1)$ random
analytic function (\ref{psi2}), and the correlations between zeros of
the random polynomial approach the ones of the random analytic
function. However, according to formula (\ref{PNas3}), there is a $1/(L+1)$
fraction of the zeros outside of the unit disk. The limiting correlations
between those, outer zeros, and between inner and outer zeros are
described by the following theorem. Let $k_n^L(z_1,\dots,z_n)$ denote
the normalized 
$n$-point correlation function (\ref{kn}) corresponding to the
parameter $L$. Indication of $L$ is important for the
theorem.   Let furthermore $k_{nN}^L(z_1,\dots,z_n)$ denote the
normalized 
$n$-point correlation function for the zeros of polynomial (\ref{psi}).

\begin{theo} \label {knN} Assume that $|z_1|,\dots,|z_m|<1$ and
$|z_{m+1}|,\dots,|z_n|>1$. Then
\begin{equation}\label{limN}
\lim_{N\to\infty} k_{nN}^L(z_1,\dots,z_n)
=k_m^L(z_1,\dots,z_m)k_{n-m}^1(z_{m+1}^{-1},\dots,z_n^{-1})\mbox{,} 
\end{equation}
so that in the limit $N\to\infty$, the inner and outer zeros become
independent, and the limiting correlations between the outer zeros coincide,
after the change of variable $z\to 1/z$ in the argument, with the
correlations between zeros of the $SU(1,1)$ random analytic function
with the parameter $L=1$.
\end{theo} 

{\it Proof.} We will first find correlations
between the outer zeros and then prove the independence of the inner
and outer zeros.

{\it Correlations between outer zeros.} Consider the random polynomial (\ref{psi}) and define
another random polynomial,
\begin{equation}\label{phi1}
\phi(z)=\frac{1}{\sqrt{C_{N+L-1}^N}}\,z^N\psi(z^{-1})\mbox{.}
\end{equation}
Observe that if $z_j$, $|z_j|>1$, is a zero of $\psi(z)$ then $z_j^{-1}$
is a zero of $\phi(z)$ and $|z_j^{-1}|<1$. Consider $\phi(z)$ in the
disk $|z|<1$. 
From (\ref{psi}),
\begin{equation}\label{phi2}
\phi(z)=a_N+\sqrt{\frac{N\dots(N+L-2)}{(N+1)\dots(N+L-1)}}\,a_{N-1}z
+\sqrt{\frac{(N-1)\dots(N+L-3)}{(N+1)\dots(N+L-1)}}\,a_{N-2}z^2+\dots
\mbox{.}
\end{equation}
As $N\to\infty$, the expressions under the radical approaches 1
from below. In addition,
we can replace $a_{N-m}$ by $a_m$, because they are the same standard
random variables. Thus, as $N\to\infty$, $\phi(z)$
approaches the random function
\begin{equation}\label{phi3}
\phi(z)=a_0+a_1z+a_2z^2+\dots\mbox{,}\qquad |z|<1
\mbox{,}
\end{equation}
which is the $SU(1,1)$ random analytic function with the parameter $L=1$.
Observe that $\phi(z)$ is a Gaussian random polynomial
and for its correlations we have formula (\ref{Knz}). From this formula, we
obtain the convergence of correlations between zeros of $\phi(z)$ as
$N\to\infty$ to the ones of the $SU(1,1)$ random analytic function
with $L=1$. 

{\it Independence.} We introduce
the random function $\eta(z)$ such that $\eta(z)=\psi(z)$ for $|z|\le 1$ and
\begin{equation}\label{xi1}
\eta(z)=\frac{1}{\sqrt{C_{N+L-1}^N}}\,z^{-N}\psi(z)\mbox{,}\qquad
|z|>1\mbox{.}
\end{equation}
Then the zeros of $\eta(z)$ and $\psi(z)$ coincide. In addition,
$\eta(z)$ is a Gaussian random field and its covariance function 
$E(\eta(z) \overline{\eta(z')})$ coincides with $E(\psi(z) \overline{\psi(z')})$ 
when $|z|,|z'|<1$,
and hence, as $N\to\infty$, it approaches the correlation function (\ref{Epsipsi}) of the
$SU(1,1)$ random analytic function with the parameter $L$.
Similarly, as we saw above in this appendix, when $|z|,|z'|>1$,
$E(\eta(z) \overline{\eta(z')})$
approaches the $SU(1,1)$ covariance function with $L=1$, if we replace
 $z,z'$ by $z^{-1},{z'}^{-1}$,
respectively. Consider now the correlation function
$E(\eta(z) \overline{\eta(z')})$ when $|z|>1$ and $|z'|<1$. From (\ref{xi1}), 
\begin{equation}\label{Exi}
E(\eta(z) \overline{\eta(z')})=\frac{1}{z^N \sqrt{C_{N+L-1}^N}}
\sum_{m=0}^N C_{L+m-1}^{m}(z\overline {z'})^m
\mbox{ .}
\end{equation}
Assume that $|zz'|\le 1$. Then we can estimate the correlation function
as follows:
\begin{equation}\label{Exi1}
\left|E(\eta(z) \overline{\eta(z')})\right|\le\frac{1}{|z|^N \sqrt{C_{N+L-1}^N}}      
\sum_{m=0}^N C_{L+m-1}^{m}\le \frac{N^L}{|z|^N}
\mbox{ .}
\end{equation}
If $|zz'|>1$, then we similarly obtain that
\begin{equation}\label{Exi2}
\left|E(\eta(z) \overline{\eta(z')})\right|=\left|\frac{{z'}^N}{\sqrt{C_{N+L-1}^N}}
\sum_{m=0}^N C_{L+m-1}^{m}(z\overline {z'})^{m-N}\right|\le {N^L}{|z'|^N}
\mbox{ .}
\end{equation}
Combining the two cases we can write
\begin{equation}\label{Exi3}
\left|E(\eta(z) \overline{\eta(z')})\right|\le N^L\left(\max\{|z|^{-1},|z'|\}\right)^N
\mbox{ .}
\end{equation}
This shows that the values of $\eta(z)$ inside and outside of the unit disk 
become independent as $N\to\infty$. Hence their zeros become independent.
Explicit estimates for the correlations between the inner and outer
zeros follow from formula (\ref{Knz}) applied to $\eta(z)$.

\vskip 5mm

{\it Acknowledgement.} Research of P.B. is partially supported 
by NSF grant \#DMS-9970625 and this support is gratefully
acknowledged.

\end{document}